\documentclass[sigconf]{acmart}

\usepackage{xcolor}
\usepackage{algorithmic}
\usepackage{graphicx}
\usepackage{textcomp}
\usepackage{subcaption}
\usepackage{xcolor}
\usepackage[utf8]{inputenc}
\usepackage{balance}
\usepackage{soul}
\usepackage{hyperref}
\usepackage{multirow}
\usepackage{fancybox}
\usepackage{adjustbox}
\usepackage{enumitem}
\usepackage{graphicx}
\usepackage{color}
\usepackage{float}
\usepackage{xcolor}
\usepackage{array}
\usepackage{amsmath}
\usepackage{xspace}
\usepackage{url}
\usepackage{pgfplots}
\usepackage{pgfplotstable}
\usepackage{tikz}
\usepackage{colortbl}
\usepackage{caption}
\usepackage{pgfplots}
\usepackage{listings}
\usepackage{color}
\usepackage{graphicx}
\definecolor{dkgreen}{rgb}{0,0.6,0}
\definecolor{gray}{rgb}{0.5,0.5,0.5}
\definecolor{mauve}{rgb}{0.58,0,0.82}
\pgfplotsset{compat=1.16}
\usepackage{pgf-pie}
\usetikzlibrary{pgfplots.statistics,calc}
\usepackage{algorithm}
\usepackage{algorithmic}
\usepackage{subcaption}
\usepackage{listings}

\usepackage{tcolorbox}

\makeatletter
\newcommand{\mybox}[1]{%
	\setbox0=\hbox{#1}%
	\setlength{\@tempdima}{\dimexpr\wd0+13pt}%
	\begin{tcolorbox}[boxrule=0.5pt, colback=gray!10, arc=4pt,
		left=6pt,right=6pt,top=6pt,bottom=6pt,boxsep=0pt]
		#1
	\end{tcolorbox}
}

\definecolor{codegreen}{rgb}{0,0.6,0}
\definecolor{codegray}{rgb}{0.5,0.5,0.5}
\definecolor{codepurple}{rgb}{0.58,0,0.82}
\definecolor{backcolour}{rgb}{0.95,0.95,0.92}

\lstdefinestyle{mystyle}{
  language=Python,
  aboveskip=3mm,
  showstringspaces=false,
  columns=flexible,
  numbers=none,
  backgroundcolor=\color{backcolour},
  commentstyle=\color{codegreen},
 keywordstyle=\color{magenta},
    numberstyle=\tiny\color{codegray},
    stringstyle=\color{codepurple},
    basicstyle=\small\ttfamily,
    breakatwhitespace=false,         
    breaklines=false,                 
    captionpos=b,                    
    keepspaces=false,                 
    numbersep=5pt,                  
    showspaces=false,                
    showstringspaces=false,
    showtabs=false,                  
    tabsize=2,
    escapeinside=``
}
\AtBeginDocument{%
  \providecommand\BibTeX{{%
    \normalfont B\kern-0.5em{\scshape i\kern-0.25em b}\kern-0.8em\TeX}}}

\setcopyright{acmcopyright}
\copyrightyear{2018}
\acmYear{2018}
\acmDOI{XXXXXXX.XXXXXXX}

\acmPrice{15.00}
\acmISBN{978-1-4503-XXXX-X/18/06}

\author{Nima Shiri Harzevili, Hung Viet Pham, Song Wang} 
    \affiliation{ 
      \institution{Lassonde School of Engineering, York University}
      \city{Toronto} 
      \country{Canada} 
    }
    \email{{nshiri, hvpham, wangsong}@yorku.ca}

\begin{document}

\title{Benchmarking Deep Learning Fuzzers}

\renewcommand{\shortauthors}{Trovato and Tobin, et al.}

\begin{abstract}
In recent years, the practice of fuzzing Deep Learning (DL) libraries has garnered significant attention in the software engineering community. Many DL fuzzers have been proposed to generate malformed inputs to test DL APIs. Although most of these fuzzers have been demonstrated to be effective in detecting bugs and outperform their respective prior work in finding more or different bugs, there remains a gap in benchmarking these DL fuzzers regarding their effectiveness against ground-truth real-world bugs in DL libraries. Since the existing comparisons among these DL fuzzers mainly focus on comparing bugs detected by them, they cannot provide a direct, in-depth evaluation of different DL fuzzers.

In this work, we set out to conduct the first ground-truth empirical evaluation of state-of-the-art DL fuzzers. Specifically, we first manually created an extensive DL bug benchmark dataset, which includes 627 real-world DL bugs from TensorFlow and PyTorch libraries reported by users between 2020 and 2022. Then we run three state-of-the-art DL fuzzers, i.e., FreeFuzz, DeepRel, and DocTer, on the benchmark by following their instructions. We find that these fuzzers are unable to detect many real bugs collected in our benchmark dataset. Specifically, most (235) of the 257 applicable bugs cannot be detected by any fuzzer.

Our systematic analysis further identifies four major, broad, and common factors that affect these fuzzers' ability to detect real bugs. These findings present opportunities to improve the performance of the fuzzers in future work. As a proof of concept, we propose a lightweight corner case generator as an extension to the three DL fuzzers, which simply covers several boundary values as well as DL-specific data types. It helps FreeFuzz, DeepRel, and DocTer detect 12, 12, and 14 more bugs, respectively, that were overlooked by the original fuzzers. Overall, this work complements prior studies on DL fuzzers with an extensive performance evaluation and provides a benchmark for future DL library fuzzing studies. Also, our proposed corner case generator proves that the fuzzers can be extended to detect more bugs by extending their internal fuzzing logic based on the insights provided in root cause analysis.

\end{abstract}

\ccsdesc[500]{Software and its engineering~Software evolution}
\ccsdesc[500]{Computing methodologies~Machine learning}
\ccsdesc[500]{Software and its engineering~Software libraries and repositories}

\keywords{Deep learning, fuzz testing, benchmarking
}

\received{20 February 2007}
\received[revised]{12 March 2009}
\received[accepted]{5 June 2009}

\maketitle

\section{Introduction}
\label{sec:intro}

In recent years, fuzz testing~\cite{takanen2018fuzzing}, which detects bugs via generating malformed inputs, has been adopted to test deep learning libraries~\cite{wei2022free, xie2022docter, deng2022fuzzing}. Specifically, DocTer~\cite{xie2022docter} is the first approach for API-level DL testing in which it extracts DL-specific input constraints from API documentation. Further, it creates rules from syntactic patterns found in API descriptions, which are then applied to a large number of API documents in popular DL libraries to extract their input parameter constraints. The extracted constraints are then used to automatically generate valid and invalid inputs to test DL API functions. FreeFuzz \cite{wei2022free} is the second step toward finding bugs in DL libraries using fuzz testing; it generates random inputs and feeds them into the target DL library to see how it reacts. Along this line, DeepRel~\cite{deng2022fuzzing} extends FreeFuzz by using test inputs from one API to test other related APIs that share similar input parameters, with the assumption that APIs with similar input parameters can share input specifications. All three DL fuzzers have been evaluated to effectively detect crashes and logical bugs in DL libraries.

Although these tools have been demonstrated to be effective in detecting bugs in DL libraries and outperforming their prior work in finding more or different bugs, there remains a gap between the evaluated result and the tools' practical effectiveness. Specifically, prior evaluation failed to completely evaluate the tools' effectiveness, bug detection scope, and limitations against ground-truth DL library bugs. This leaves an unanswered question: \textit{How effectively and thoroughly can these DL fuzzers find bugs in practice?}

\begin{table*}[t!]
\caption{Characteristics of DL fuzzers used in this study.} 
\vspace{-0.1in}
\centering
\renewcommand{\arraystretch}{1}
\resizebox{1.8\columnwidth}{!}{%
\begin{tabular}{llccccccc}
\toprule
\multirow{2}{*}{Fuzzer} & \multicolumn{4}{c}{Mutation Strategy} & \multicolumn{4}{c}{Oracles}      \\
\cmidrule(lr){2-5} \cmidrule(lr){6-9}
& \multirow{1}{*}{Boundary-input} & \multirow{1}{*}{Shape mutation} & \multirow{1}{*}{Value/Type mutation} & \multirow{1}{*}{Valid value} & \multirow{1}{*}{Crash} & \multirow{1}{*}{CPU/GPU} & \multirow{1}{*}{Performance} & \multirow{1}{*}{Value Inconsistency} \\ \midrule
FreeFuzz                & Random                          & $\checkmark$                               & $\checkmark$                                  & $\times$                            &$\checkmark$                      & $\checkmark$                    & $\checkmark$                         & $\times$                           \\
DeepRel                 & Random                          & $\checkmark$                             & $\checkmark$                                    & $\times$                          & $\checkmark$                      & $\times$                    &$\times$                          & $\checkmark$                              \\
DocTer                  & Rule-based                      & $\checkmark$                               & $\checkmark$                                    & $\checkmark$                           &$\checkmark$                     & $\times$                     & $\times$                         & $\times$                           \\
\bottomrule
\end{tabular}
}
\label{tbl:toolcomparison}
\end{table*}

To our knowledge, no effort exists in the literature to evaluate DL fuzzers for detecting real bugs from an extensive ground-truth bug dataset. In this paper, we conduct the first extensive and systematic empirical study to compare three well-known DL fuzzers, including DocTer \cite{xie2022docter}, FreeFuzz \cite{wei2022free}, DeepRel \cite{deng2022fuzzing} on two popular and open-source DL libraries (TensorFlow and PyTorch). The first important step is to collect a ground-truth benchmark with real-world bugs from DL libraries. Specifically, we manually created an extensive DL bug benchmark dataset that includes 627 real-world DL bugs from TensorFlow and PyTorch reported between 2020 and 2022. To collect the bugs, we automatically extracted GitHub bugs using a keyword-matching approach~\cite{zhou2017automated} from TensorFlow and PyTorch repositories. Since the automatic extraction may introduce false positives, we further conducted a manual analysis process to verify each bug collected. In the second step, we rigorously run the three DL fuzzers by following their instructions on the corresponding release versions of the collected bugs.

Our analysis explores the performance of each DL fuzzers (Section~\ref{sec:result}) from the following three aspects: the number of bugs it can theoretically find (i.e., scope), the number of bugs it actually detects (i.e., effectiveness), and how many inputs it generates to detect bugs on average (i.e., efficiency). Finally, we present a detailed quantitative and qualitative analysis of the detected bugs for each DL fuzzer and examine these fuzzers’ implementations to identify the challenges that existing tools face and the factors that affect their bug detection abilities. In this work, we investigate the following research questions:

\vspace{4pt}
\noindent \textbf{RQ1:} \textit{What is the distribution of bugs extracted from DL libraries?}

\vspace{4pt}
\noindent \textbf{RQ2:} \textit{How effectively and thoroughly can these fuzzers detect real-world DL bugs?}

\vspace{4pt}
\noindent \textbf{RQ3:} \textit{What are the characteristics of detected DL bugs?}

\vspace{4pt}
\noindent \textbf{RQ4:} \textit{What are the reasons for missing detecting bugs?}

Overall, we find that these fuzzers miss detecting a large number of the real bugs in our benchmark dataset, i.e., 235 out of the 257 applicable bugs cannot be detected by any fuzzer. Our analysis reveals that DocTer is effective in detecting crash bugs due to its strong crash oracle, while FreeFuzz and DeepRel are capable of detecting crashes and a limited number of logical bugs since they support both crash and precision oracles. 
Finally, we discuss implications that can improve future DL library fuzzers by analyzing the root cause of missing bugs that are outside of the detection scope of the studied fuzzers. As a proof of concept, we propose a lightweight corner case generator as an extension to the three DL fuzzers, which simply covers several boundary values as well as DL-specific data types. It helps FreeFuzz, DeepRel, and DocTer detect 12, 12, and 14 new bugs, respectively, that were overlooked by the original fuzzers. Overall, we believe this work complements prior studies on DL fuzzers with an extensive performance evaluation and provides a benchmark for future DL library fuzzing studies.

This paper makes the following contributions:

\begin{itemize}
    \item  We take the first step to conduct an empirical study against real-world DL library bugs to evaluate state-of-the-art DL fuzzers, which provides a complementary perspective on prior studies regarding the comparison among DL fuzzers.
    
    \item We create the first benchmarking dataset for evaluating DL fuzzers rigorously, which includes 627 reproducible bugs from TensorFlow and PyTorch. 
    
    \item We conduct an in-depth quantitative and qualitative evaluation of DL fuzzers and present findings regarding the scope, effectiveness, and efficiency of these DL fuzzers. 
    
    \item We release the dataset and source code of our experiments to help other researchers replicate and extend our study\footnote{https://anonymous.4open.science/r/Benchmarking-DL-Fuzzers-AED2/README.md}.
\end{itemize}

The rest of this paper is organized as follows. 
Section~\ref{sec:approach} describes the methodology of our approach. 
Section~\ref{sec:experiment} and Section~\ref{sec:result} respectively show the experimental setup and the evaluation results. 
Section~\ref{sec:discussion} discusses the threats to the validity of this work. 
Section~\ref{sec:corner} shows our corner case generator for improving DL fuzzers. 
Section~\ref{sec:related} presents the related studies. 
Section~\ref{sec:threats} discuss the threats to the validity of this work. 
Section~\ref{sec:conclusion} concludes this paper.

\section{Approach}
\label{sec:approach}
{In this section, we present our approach to constructing a benchmark to evaluate the effectiveness of the studied DL fuzzers. First, we explain the process of selecting DL fuzzers for our benchmark analysis (Section~\ref{sec:step1}). We then discuss the selection of subject DL libraries (Section~\ref{sec:step2}). Finally, we explain how we gathered real-world bugs using a combination of automatic filtering and manual analysis (Section~\ref{sec:step3}).}


\subsection{Selection of DL fuzzers}
\label{sec:step1}

The focus of our research is on the examination of three state-of-the-art DL fuzzers: FreeFuzz \cite{wei2022free}, DeepRel \cite{deng2022fuzzing}, and DocTer \cite{xie2022docter}, which are currently being developed and freely available to the public. 
Even though there are multiple fuzzing tools proposed both in the academia~\cite{ba2022efficient, kallingal2022fuzzeraid, lee2022fuzzle, fu2022griffin, yu2022htfuzz} and the industry \cite{ AFL, libFuzzer, Fuzzilli}, in this paper, we give particular attention to fuzzers that concentrate on DL library fuzzing. There are a few reasons why we specifically concentrate on API-level fuzzing of high-level DL APIs written in Python. First, DL libraries are commonly used via their Python APIs from the front end, while the actual computations are performed in the DL backend (which is mainly programmed in C/C++). While some general fuzzers use C/C++ APIs for fuzzing, testing DL libraries in their typical usage (i.e., via Python APIs) is the most effective way to detect bugs that are important to users. Second, DL libraries differ significantly from traditional software systems \cite{simhambhatla2019self,harzevili2022characterizing} primarily due to their use of tensors--a unique data structure in which all computations occur at the tensor level. We chose to focus on DL fuzzers because they can simulate the tensor data structure for fuzzing, whereas general fuzzers are only capable of modeling preemptive data structures.

Table \ref{tbl:toolcomparison} outlines the key characteristics of the three fuzzers analyzed in this paper. We performed a manual analysis of the source code and documentation for each fuzzing tool to assess its capabilities and limitations. We characterize DL fuzzers mainly based on their mutation strategies and types of oracles. This comprehensive analysis allows us to compare the effectiveness of each fuzzer.


\noindent \textbf{DocTer~\cite{xie2022docter}} {is a fuzzing tool that uses two major techniques for testing DL APIs:
\textit{DL-specific constraint extraction} and \textit{DL-specific input generation}. Initially, DocTer extracted input constraints specific to DL libraries from API documentation. It generates rules from syntactic patterns found in API descriptions and applies them to a large number of API documents in popular DL libraries to extract their input parameter constraints. The constraints are then used to guide its fuzzer to generate conform, violate, and boundary inputs to DL APIs. Once the test cases are generated, DocTer executes them and performs stack trace analysis to look for crash bugs such as \textit{Segmentation Fault}, \textit{Bus Errors}, \textit{Aborts}, and \textit{Floating point exceptions}.


\noindent \textbf{FreeFuzz~\cite{wei2022free}}
{is an API-level fuzz testing tool that performs test case generation using random value and type mutation. It also supports boundary input values to generate test cases that find edge cases in DL libraries. However, the boundary input generation is random, unlike DocTer, where rules are used to guide the boundary input generation. FreeFuzz uses three types of oracles: crash, cuda, and precision. Like DocTer, FreeFuzz also uses the same crash oracle, i.e., by analyzing the log message produced by the test case. In terms of Cuda Oracle, FreeFuzz runs the test case under two configurations: CPU and GPU. Then, using differential testing, it compares the output of APIs under two settings. Inconsistent outputs between these two settings indicate a potential bug. Regarding precision oracles, FreeFuzz considers the metamorphic relation between two APIs with the same input specification, e.g., Tensors with the same data type and dimensions, to determine which one is executing faster compared to the other one.


\noindent \textbf{DeepRel~\cite{deng2022fuzzing}} {extends FreeFuzz by using test inputs from one API to test other related APIs that share similar input parameters. It hypothesizes that APIs with similar input parameters can share input specifications. Like FreeFuzz, it also supports type and value mutation of preemptive data structures as well as tensor data structures. Its major difference when compared to FreeFuzz and DocTer is the oracle where DeepRel compares two semantically related APIs in terms of equivalence value and status.}

\subsection{DL libraries selection}
\label{sec:step2}

In this study, we chose TensorFlow \cite{abadi2016tensorflow} and PyTorch \cite{paszke2017automatic} which are both very popular DL frameworks that have been widely used in the literature~\cite{lemon, pham2019cradle, cao2022mvd, li2022dear} for two main reasons. First, there are numerous usages of TensorFlow and PyTorch in various application domains, including image classification \cite{algan2021image, mahdisoltani2018fine}, big data analysis \cite{ertam2017data}, pattern recognition \cite{lv2022semi}, self-driving \cite{simhambhatla2019self, ramos2017detecting, kulkarni2018traffic} and Natural Language Processing \cite{minaee2017automatic, athreya2021template, roy2021deep}. 
Second, the three benchmark fuzzers used in this paper have used TensorFlow and PyTorch as their target DL libraries for fuzzing.


\subsection{Automated collection of real-world DL bugs}
\label{sec:step3}

{This study aims to evaluate the efficacy of three state-of-the-art DL fuzzers in practice. To this end, we first need to collect a set of real-world bugs.
In this paper, we use the term ``bug'' in a general sense to refer to any kind of software defect, including security vulnerabilities, logical bugs, and performance bugs. 
We first collect bugs from the GitHub repository of our subject libraries (i.e., TensorFlow and PyTorch)\footnote{The GitHub repository can be considered a reliable source of bugs since the users often report bugs as issues}. 
Since all the studied fuzzers require APIs with input parameters from DL libraries to conduct fuzzing testing, for each bug in our dataset, we identify and collect the corresponding buggy APIs and the library release in which the bug can be replicated. }

We adopt the following two steps for automatic data collection: 1) We first filter bugs with the labels \textit{bug} and \textit{bug-fix} and 2) for each bug collected in step 1), we automatically analyze the title, body, and comments and search for bug-related keywords using the approach proposed in \cite{zhou2017automated}.  Specifically, we use the following keywords for different types of bugs:

\noindent \textbf{Security vulnerabilities}: \textit{buffer overflow}, \textit{integer overflow}, \textit{cross-site scripting}, \textit{remote code execution}, \textit{memory leak}, \textit{race condition}, \textit{heap buffer overflow}, and \textit{null pointer dereference}\footnote{Please note that due to brevity, we do not include all security keywords in the manuscript. We list all keywords in the GitHub repository of this paper.}.

\noindent \textbf{Logical errors}: \textit{wrong result}, \textit{unexpected output}, \textit{incorrect calculation}, \textit{inconsistent behavior}, \textit{unexpected behavior}, \textit{incorrect logic}, \textit{wrong calculation}, \textit{logic error}.

\noindent \textbf{Performance errors}: \textit{slow}, \textit{high CPU usage}, \textit{high memory usage}, \textit{poor performance}, \textit{slow response time}, \textit{performance bottleneck}, \textit{performance optimization}, \textit{resource usage}.

At the end of this automated process, 2,459 issues from PyTorch and 334 issues from TensorFlow were collected for our manual validation. The reason behind this unbalanced number of reported issues is that PyTorch has a higher number of reported issues compared to TensorFlow. Additionally, the TensorFlow community also reports the security weaknesses in their security advisory\footnote{https://github.com/tensorflow/tensorflow/security/advisories}. That is why we also leveraged the security records reported in the advisory. Hence, we collect 179 security records that have already been analyzed and confirmed by the community as security vulnerabilities; hence, we exclude them from our manual analysis. While this automatic mining of GitHub repositories proved efficient in gathering a substantial number of issues, it also led to the inclusion of numerous unrelated or non-bug-related issues. To ensure the integrity of our manual analysis and focus solely on potential bugs, we chose to manually analyze the collected issues gathered from TensorFlow and PyTorch repositories. This thorough manual analysis helped us identify and validate actual bugs effectively.}

\subsection{Manual validation to create our benchmark dataset}
\subsubsection{Manual Analysis Criteria}
Once we automatically collected the related bugs from the GitHub repository of TensorFlow and PyTorch, we performed manual analysis to filter false-positive instances. Here are the definitions of bugs that we used to guide our manual analysis: 

\noindent \textbf{Security vulnerability}: A security vulnerability or weakness is a flaw in the source code of any general software or DL library that external attackers can exploit to take control of the software \cite{chess2007secure}. In our manual analysis, we are looking for weaknesses that are triggered by feeding malicious inputs to DL APIs. 

\noindent \textbf{Logical bug}: A logical bug \cite{tan2014bug} in a DL library refers to an error in the code that is not related to syntax or other obvious mistakes but rather to a flaw in the reasoning or logic behind the algorithm. In our benchmark, we include logical bugs that are triggered by malicious DL API inputs. 

\noindent \textbf{Performance bug}: A performance bug \cite{han2016empirical} in a DL library refers to a bug that degrades the efficiency or speed of the DL model's training or inference process. Performance bugs can occur for a variety of reasons, such as inefficient algorithm design, suboptimal hardware utilization, poor memory management, or excessive I/O operations. These bugs can negatively impact the scalability and usability of the DL model, making it harder to train and deploy at scale. In this study, we look for performance bugs that are caused by malicious inputs to DL APIs, e.g., feeding a very large input to trigger an Out of Memory (OOM) error.

For each bug, we examined its title, bug body, comments, linked PRs, and commits to determine whether the bug is related or not. We exclude the following types of unrelated bugs:

\begin{itemize}
    \item Bugs that are specific to certain platforms, such as Windows, Apple M chips, Android, or IOS.
    \item Build and configuration bugs.
    \item Bugs in the backend, no high-level API is involved for triggering the bug.
    \item Bugs from external libraries such as torchvision.
    \item Bugs that do not require input parameters to be triggered.
\end{itemize}


For each collected bug, we retrieved the bug type, and release information for the involved APIs (the release in which the bugs can be replicated). 

For each bug, the affected and patched releases are specified. In our experiments, we randomly chose one affected release that has not yet been patched. Users may report numerous releases affected by the bug, but we opt to consider only one affected release per bug when running the fuzzers. We also limit the data collection interval from 2020 to July 2022 (the time when the first fuzzer, i.e., DocTer~\cite{xie2022docter}, was released) to avoid bugs that were already identified and reported by the three studied fuzzers and maintain a fair evaluation.

\subsubsection{Manual Analysis Procedure} 
Starting with the 2,459 and 334 automatically extracted bugs for PyTorch and TensorFlow repositories, respectively, the authors conducted a manual analysis with two rounds:

\noindent \textbf{Round 1}: 
Three authors independently reviewed PyTorch and TensorFlow and collected GitHub bugs. The authors extracted multiple pieces of information from each bug, i.e., the bug type, the involved APIs, and the affected release. Once the information is extracted, the authors cross-check the labeled bugs to mark possible disagreements (e.g., lack of consensus on the bug type). In this step, the disagreement rates were 27.3\% and 19.1\% for PyTorch and TensorFlow, respectively. {In the end, 429 and 58 bugs (including disagreements) from PyTorch and TensorFlow are kept in the dataset.}

\noindent \textbf{Round 2}: 
In this round, all authors were involved in the manual analysis of the 429 and 58 bugs in PyTorch and TensorFlow from Round 1, and disagreements were resolved with group discussions. The rate of disagreements at the end of this round was 0.02 and 0.01 for PyTorch and TensorFlow, respectively. At the end of the round, we discarded any bug on which the authors could not reach a consensus. Finally, 393 and 55 bugs for PyTorch and TensorFlow are left in the dataset. Note that, in addition to these 55 TensorFlow bugs, we also collected 179 security records from the TensorFlow security advisories, resulting in 234 TensorFlow bugs in our benchmark.

\section{Experiment Setup}
\label{sec:experiment}

\subsection{Experiment Data}

Table \ref{tbl:data} shows the number of collected bugs and unique APIs in our benchmark. The total number of bugs (\# Bugs) we could get after our manual analysis is 393 and 234 for PyTorch and TensorFlow. Since the internal database of the fuzzers may not cover all APIs we collected, We consider a bug applicable if its buggy DL API exists in the internal database of the target fuzzer. Column \textit{\# Applicable Bugs} shows the total number of applicable bugs that we collected which are also present in the internal database of the employed fuzzers, while column \textit{\# Unique Applicable APIs} shows the total number of corresponding applicable unique APIs.


\begin{table}[t!]
\centering
\caption{Number of collected bugs
}
\vspace{-0.1in}
\renewcommand{\arraystretch}{1}
\resizebox{1.01\columnwidth}{!}{%
\label{tbl:data}
\begin{tabular}{lcccc}
\toprule
Library    & \# 
Bugs & \# Unique APIs
& \#  Applicable Bugs 
& \# Unique Applicable APIs 
\\ 
\midrule
PyTorch    & 393                    & 278            &     201          &   152          \\
TensorFlow & 234                    & 194            &     56           &   58         \\ \hline
Total      & 627                    & 472            &     257          &   210       \\
\bottomrule
\end{tabular}}
\end{table}

\subsection{Configuring DL fuzzers}
{In this section, we describe the steps we took to set up the DL fuzzers used in our benchmark analysis. It should be noted that we followed the provided instructions in the replication documentation of each DL fuzzer.}

\noindent \textbf{FreeFuzz~\cite{wei2022free}}:
To run FreeFuzz, we enabled crash, cuda, and precision oracles. We also set the float difference cut-off value as its default value of $1e-2$. We also set \textit{max\_time\_bound} and \textit{time\_thresold} to the default values of $10$ and $1e-3$ respectively. Regarding the mutation strategies, we enabled value mutation, type mutation, and database mutation. We also set the number of times FreeFuzz performs testing for each API to 500.

\noindent \textbf{DeepRel~\cite{deng2022fuzzing}}:
Compared to FreeFuzz, there are three more hyperparameters to be set for DeepRel. 
We set the test number (i.e., the number of times each API is executed to generate a test case) to 500 (meaning that for each API, 500 test cases are generated with different inputs), \textit{top\_k} parameter to 3, and the number of iterations to 1 (as instructed).
Note that for both FreeFuzz and DeepRel, we use the same internal database for both TensorFlow and PyTorch libraries. 

\noindent \textbf{DocTer~\cite{xie2022docter}}: There are multiple DocTer hyperparameters that we have to configure. The most important parameter is the fuzzing mode which we set as \textit{violate}.  
We also set \textit{fuzz\_optional\_p=0.2}, \textit{mutate\_p=0.4}, and \textit{timeout=10}. We fuzz each API 500 times as suggested and use DocTer's own rules and internal database for fuzzing.

We perform the experiment on a machine equipped with Intel(R) Core(TM) i7-10700F CPU @ 2.90GHz, NVIDIA GTX 1660 Ti GPU, 16GB of RAM, Ubuntu 22.04, Python 3.6 to 3.10 for different releases of PyTorch and TensorFlow. Using our machine configuration, the time cost of running the fuzzers is shown in Table. \ref{tbl:timeCost}. 

\begin{table}[h]
\centering
\caption{Time cost of running each fuzzer (in terms of CPU hours)}
\vspace{-0.1in}
\begin{tabular}{lccc}
\toprule
Tool     & PyTorch & TensorFlow & Total \\
\midrule
FreeFuzz & 20     & 30        & 50   \\
DeepRel  & 30     & 40        & 70   \\
DocTer   & 24     & 35        & 59   \\
\bottomrule
\end{tabular}
\label{tbl:timeCost}
\end{table}


\subsection{Evaluation Metrics}
In this study, we then use the following metrics to measure the performance of each DL fuzzer based on the number of detected bugs:

\noindent {\textbf{Number of generated test cases}} is a typical metric to assess the effectiveness of the studied DL fuzzers. This metric counts the number of distinct test cases generated by each fuzzer.

\noindent {\textbf{Number of true bugs detected}}
{determines the number of real-world bugs discovered by a fuzzer. To calculate this metric, we performed a manual cross-check between every potential bug in each PyTorch and TensorFlow release with bugs in our benchmark. This involves manually matching the log messages or stack traces of the generated test cases by a fuzzer under each release with the log messages or stack traces reported in real-world bugs. We considered a bug detected by a fuzzer to be true if there was a match between the log message or stack traces of any generated test case and the bug.}








\section{Results and Analysis}
\label{sec:result}

\subsection{RQ1: Distribution of Bugs Studied}
\label{sec:answer_rq1}
\noindent \textbf{Approach:} In this RQ, we investigate the distribution of bug types extracted by mining issues from the GitHub repositories of PyTorch and TensorFlow. As the basic information of each bug has already been collected in our data collection (Section~\ref{sec:step3}), we directly analyze our bug dataset for answering this RQ. 

\pgfplotsset{width=7cm,compat=1.8}

\begin{figure}[t!]
\centering
\renewcommand{\arraystretch}{1}
\resizebox{1.05\columnwidth}{!}{%
\begin{tikzpicture}
  \centering
  \begin{axis}[
        ybar, axis on top,
        height=3.5cm, width=13.5cm,
        bar width=0.4cm,
        ymajorgrids, tick align=inside,
        major grid style={draw=white},
        enlarge y limits={value=.1,upper},
        ymin=0, ymax=120,
        axis x line*=bottom,
        axis y line*=left,
        y axis line style={opacity=0},
        tickwidth=0pt,
        enlarge x limits=true,
        legend style={
            at={(0.8,1)},
            anchor=north,
            legend columns=1,
            /tikz/every even column/.append style={column sep=0.5cm}
        },
        ylabel={\# of bugs}, 
        xticklabel style={rotate=25,anchor=east},
        symbolic x coords={
Runtime Error,
Nullptr Dereference,
Division By Zero,
Inconsistent Results,
Out of Bound Read,
Internal Assertion Failure,
Segfault,
Heap Buffer Overflow,
Performance Bug,
Type Error,
Crash,
Memory Leak
           },
       xtick=data,
       nodes near coords={
        \pgfmathprintnumber[precision=0]{\pgfplotspointmeta}
       }
    ]
    \addplot [draw=none, fill=blue!30] coordinates {
(Runtime Error, 119)
(Nullptr Dereference, 0)
(Division By Zero, 0)
(Inconsistent Results, 86)
(Out of Bound Read, 0)
(Internal Assertion Failure, 41)
(Segfault, 27)
(Heap Buffer Overflow, 0)
(Performance Bug, 22 )
(Type Error, 12)
(Crash, 11)
(Memory Leak, 10)
};
   \addplot [draw=none,fill=red!30] coordinates {
(Runtime Error, 0)
(Nullptr Dereference, 36)
(Division By Zero, 30)
(Inconsistent Results, 24)
(Out of Bound Read, 22)
(Internal Assertion Failure, 49)
(Segfault, 11)
(Heap Buffer Overflow, 15)
(Performance Bug, 0)
(Type Error, 4)
(Crash, 6)
(Memory Leak, 3)
   };
    \legend{PyTorch,TensorFlow}
  \end{axis}
  \end{tikzpicture}}
  \vspace{-0.1in}
\caption{Top bug types in PyTorch and TensorFlow libraries}
\label{fig:bugdist}
  \end{figure}
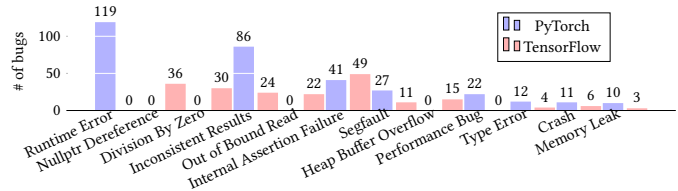

\noindent \textbf{Result:} Figure \ref{fig:bugdist} shows the distribution of bug types in the PyTorch and TensorFlow libraries. The x-axis represents the different bug types, and the y-axis shows the number of issues reported for each bug type. The blue bars show the PyTorch distribution, while the red ones show the distribution of bugs in TensorFlow. \textit{Runtime error} is the most frequent bug type, accounting for 119 issues in the PyTorch library. \textit{Inconsistent result}, \textit{Internal Assertion failure}, and \textit{Segmentation fault} are also relatively high, indicating a significant number of issues reported for these bug types. Regarding the TensorFlow library, we can observe that \textit{Internal Assertion failure} is the most common type of bug in the TensorFlow library, with a total of 49 reported issues. \textit{Nullptr dereference} is the second most common type, with 36 reported issues, followed by \textit{Division by zero} with 30 issues.

\mybox{\textbf{Answer to RQ1:} \textit{Runtime error} is the most frequent bug type, with 119 issues reported in the PyTorch library. \textit{Internal Assertion Failure} is the most common type of bug with 49 reported issues in the TensorFlow library, followed by \textit{Nullptr dereference} with 36 issues and \textit{Division by zero} with 30 issues.}

\subsection{RQ2: Performance of Fuzzers}
\label{sec:answer_rq2}

\begin{table*}[h]
\centering
\caption{Results of FreeFuzz, DeepRel, and DocTer on different releases of PyTorch. We also show the total number of generated test cases. \# Hit shows the number of bugs detected by a fuzzer on a release that matches the bugs in our benchmark dataset.}
\vspace{-0.1in}
\begin{adjustbox}{width=0.72\textwidth}
\begin{tabular}{lcccccccccc}

\hline
\multicolumn{2}{l}{PyTorch}      & \multicolumn{3}{c}{FreeFuzz}                 & \multicolumn{3}{c}{DeepRel}                  & \multicolumn{3}{c}{DocTer}                   \\
\cmidrule(lr){3-5} \cmidrule(lr){6-8} \cmidrule(lr){9-11}
Release & Total bugs & Applicable bugs & \# test cases & \# Hit & Applicable bugs & \# test cases & \# Hit & Applicable bugs & \# test cases & \# Hit \\
\midrule    
v1.0.0  & 3                      & 1                & -             &  -         & 1                &     -              &   -         &   2               & -             &   -        \\
v1.4.0  & 1                      & 1                & 295           &  0         & 1                &     2240           &  0         &   -               & 36             &   0        \\
v1.5.0  & 2                      & 2                & 5033          &  0         & 2                &     2896           &   0        & 1                & 13             &      0     \\
v1.6.0  & 2                      & 2                & 1858          &   0        & 2                &     2880           &   0        & 1                & 342             &    0       \\
v1.7.0  & 26                     & 9                &   0           &   -        & 9                & 3142               &   0       & 8                & 352             &   \cellcolor[HTML]{9AFF99}1        \\
v1.7.1  & 31                     & 11               & 2826          &   0        & 11               &    3123            &   0        & 11               & 357             &    0       \\
v1.8.0  & 23                     & 6                & 5371          &    0       & 6                &    5563           &  \cellcolor[HTML]{9AFF99}1         & 7                & 355             &     0      \\
v1.8.1  & 21                     & 9               & 70            &    0       & 9               &     5514          &  0        & 11               & 345             &     0      \\
v1.8.2  & 2                      & -                &  -            &  -         & -                &     -              &   -        & 1                & 345             &   0        \\
v1.9.0  & 44                     & 14               & 79            &    \cellcolor[HTML]{9AFF99}1       & 14               &      5529         & 0          & 9                & 393             &   0        \\
v1.9.1  & 15                     & 6                & 59            &    0       & 6                &      5497         & 0          & 6                & 397             &  0         \\
v1.10.0 & 45                     & 17               & 7196          &     0      & 17               &   5939            & \cellcolor[HTML]{9AFF99}2          & 14               & 0             &     0      \\
v1.10.1 & 68                     & 42               & 7313          &    \cellcolor[HTML]{9AFF99}7       & 42               &   5929            & 4           & 25               & 355             &      0     \\
v1.11.0 & 96                     & 50               &  1796         &   0        & 50               &     0              &  -         & 33               & 391            &  0         \\
v1.12.0 & 11                     & 5                &  7367         &    0       & 5                &    5969           &  0         & 4                & 401            &    0       \\
v1.13.0 & 3                      & 1                &  6740         &   0        & 1                &     5810          &   0        & 1                & 379             &       0    \\
\hline
Total   & 393                    & 176              &    46003           &    8       & 176              &   60031            &  7         & 134              &   4461            &  1         \\
\hline
\end{tabular}
\end{adjustbox}
\label{tbl:maintorch}
\end{table*}

\begin{table*}[h]
\centering
\caption{Results of FreeFuzz, DeepRel, and DocTer on different releases of TensorFlow library.} 
\vspace{-0.1in}
\begin{adjustbox}{width=0.72\textwidth}
\begin{tabular}{lcccccccccc}
\hline
\multicolumn{2}{l}{TensorFlow} & \multicolumn{3}{c}{FreeFuzz}                 & \multicolumn{3}{c}{DeepRel}             & \multicolumn{3}{c}{DocTer}                   \\
\cmidrule(lr){3-5} \cmidrule(lr){6-8} \cmidrule(lr){9-11}
Release         & Total bugs        & Applicable bugs & \# test cases & \# Hit & Applicable bugs & \# test cases & \# Hit & Applicable bugs & \# Test cases & \# Hit \\
\midrule
v2.0.0          & 1            & -                &   -            &   -        & -                &   -       &    -        & -                & -             & -         \\
v2.2.0          & 1            & -                &   -            &   -        & -                &  -        &     -       & -                & -             & -         \\
v2.3.0          & 5            & 1                & 2414          &    0        & 1                &    5451      &    0      & 1                & 517             & 0         \\
v2.3.1          & 1            & -                &   -            &   -        & -                &   -       &      -      & -                & -             & -         \\
v2.4.0          & 126          & 13               & 3482          &    0        & 13               &  5554        &  \cellcolor[HTML]{9AFF99}1        & 6                & 436            & \cellcolor[HTML]{9AFF99}2         \\
v2.5.0          & 1            & -                &  -             &   -        & -                &  -        &    -        &  -                & -             & -         \\
v2.6.0          & 31           & 12               & 38            &   \cellcolor[HTML]{9AFF99}1         & 12               &  0        &     0        & 12               & 0             & -         \\
v2.7.0          & 16           & 2                & 5649         &    \cellcolor[HTML]{9AFF99}1         & 2                &   2843       &    0        & 2                & 124             & 0         \\
v2.8.0          & 22           & 4                & 5480         &    0         & 4                &    2850      &  0         & 6                & 137            & \cellcolor[HTML]{9AFF99}1         \\
v2.9.0          & 14           & 6                & 4817          &    0        & 6                &    2830      &  0        & 8                & 6             & 0         \\
v2.9.1          & 1            & -                &  -             &    -       & -                &   -       &    -        & -                & -             & -         \\
v2.10.0         & 2            & -                &  -             &    -       & -                &   -       &     -       & 1                & 0             & -         \\
v2.11.0         & 13           & 5                & 5612          &     0       & 5                &   2682       &  0       & 5                & 0             & -         \\
\hline
Total           & 234          & 43               &      27492         &    2       & 43               &   22210       &  1         & 41               & 1220            & 3         \\
\hline
\end{tabular}
\end{adjustbox}
\label{tbl:maintf}
\end{table*}

\textbf{Approach:}
To address this RQ, we first create an anaconda environment for the corresponding release of each applicable bug (either PyTorch or TensorFlow) with CUDA and the necessary dependencies of fuzzers enabled. After that, we run fuzzer to generate tests for the involved APIs of the bug. We ran all the test cases generated by the fuzzer and stored the stack trace logs and runtime error messages. Finally, we manually compared the reported bugs by fuzzers with the ground-truth real bugs to determine if the reported bugs overlapped with the real bugs.

\textbf{Result:} Table \ref{tbl:maintorch} and Table \ref{tbl:maintf} present the detection results of three tools, i.e., FreeFuzz, DeepRel, and DocTer, on various releases of the PyTorch and TensorFlow libraries. The table is divided into three broad columns and two independent columns. The first column shows the release number of a DL library, and the second column shows the total number of issues collected that affect the release shown in the first column. There are three sub-columns under each fuzzer that indicate the total number of applicable issues for a specific fuzzer, the number of test cases generated by the fuzzer on each release, and finally, the last sub-column presents the number of {hits, i.e., the number of detected bugs by the fuzzer that match the real-world applicable bug in our benchmark dataset}.

{We report the hit rate of the fuzzers on the different releases of PyTorch and TensorFlow based on the applicable bugs for each release.Then the hit rate of a fuzzer on a specific library refers to the total number of bugs that can be detected by the fuzzer on all releases of the target library. We can observe that FreeFuzz detected the most bugs with a hit rate of 4.55\%, followed by DeepRel, i.e., the hit rate is 3.9\% in the PyTorch library. DocTer is the weakest fuzzer in terms of hit rate in the PyTorch library, with only 0.7\%. In TensorFlow, DocTer is the best fuzzer, detecting 7.3\% of real-world bugs, followed by FreeFuzz and DeepRel, which detected 4.6\% and 2\% respectively.}

On different releases of the PyTorch library, FreeFuzz and DeepRel generate 46,003 and 60,031 test cases, respectively, while DocTer only generates 4,461 test cases. The pattern is also present in the TensorFlow library, where FreeFuzz and DeepRel can jointly create 49,702 test cases, whereas DocTer generates only 1,220 test cases. The large number of test cases generated by FreeFuzz and DeepRel is due to the fact that they used more oracles and a larger input space than DocTer.

\mybox{\textbf{Answer to RQ2:} Out of the 627 bugs in the curated dataset, 34.92\% of are applicable bugs for FreeFuzz and DeepRel, while 23.12\% of bugs applicable for DocTer. FreeFuzz has the highest hit rate of 4.55\%, followed by DeepRel with 3.9\% in the PyTorch library.  In contrast, in the TensorFlow library, DocTer is the best fuzzer with a hit rate of 7.3\%.} 


\subsection{RQ3: Characteristics of Detected Bugs} 
\label{sec:answer_rq3}
\noindent \textbf{Approach:}
To address this RQ, We further manually check the detected bugs by each fuzzer and summarized them in Figure \ref{fig:detectedbugdist}. 


\noindent \textbf{Result:}
Figure \ref{fig:detectedbugdist} shows the summary of bugs detected by the studied fuzzers. \textit{RuntimeError} is the most common root cause among the bugs detected by all three fuzzers, with FreeFuzz and DeepRel detecting 5 and 4 bugs, respectively, while DocTer did not detect any "RuntimeError" bugs. \textit{InconsistentResults} is the second most common root cause, with both FreeFuzz and DeepRel detecting two bugs, while DocTer did not detect any. DocTer could detect the fewest bugs since it is only equipped with a crash oracle. Overall, the figure clearly illustrates the different bug detection capabilities of the three fuzzers, with FreeFuzz and DeepRel being more versatile due to their ability to detect bugs across multiple root causes, while DocTer focuses solely on crash bugs due to having a crash oracle. The results indicate that the choice of fuzzing tool can significantly impact the types of bugs that can be discovered during the testing process.

\mybox{\textbf{Answer to RQ3:} The three DL fuzzers can detect a total of 22 real-world bugs, with \textit{Runtime Error} being the most frequently detected bug type, accounting for 9 bugs. The fuzzers are able to detect 22 real-world bugs since the input specification of the buggy APIs is within the scope of the fuzzers input boundaries.}

\pgfplotstableread{
Tool FreeFuzz DeepRel DocTer
RuntimeError 5 4 0
InconsistentResults 2 2 0
Abort 0 0 2
IndexError 2 0 0
Crash 0 1 1
OOM 0 1 1
Segfault 1 0 0
    }\testdata
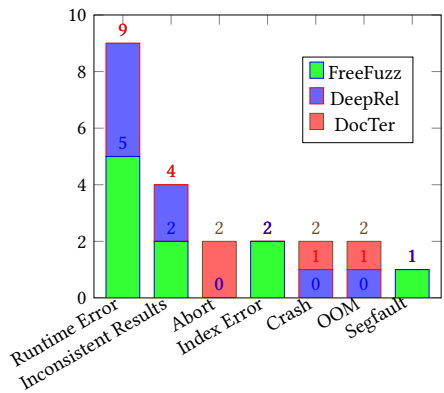
\begin{figure}[t!]
     \centering
  \renewcommand{\arraystretch}{1}
\resizebox{0.7\columnwidth}{!}{%
    \begin{tikzpicture}
    \begin{axis}[
        ybar stacked,
    	bar width=15pt,
     height = 6cm,
        ymin=0,
        ymax=10,
        xtick=data,
        legend style={at={(0.6,0.7)},anchor=west},
        xticklabels={{Runtime Error}, {Inconsistent Results}, {Abort}, {Index Error}, {Crash}, {OOM}, {Segfault}},
        xticklabel style={rotate=30,anchor=east},
    ]
    \addplot + [ nodes near coords, fill=green!80]      table    [y=FreeFuzz, meta=Tool, x expr=\coordindex] {\testdata};
    \addplot + [nodes near coords, fill=blue!60]         table    [y=DeepRel, meta=Tool, x expr=\coordindex] {\testdata};
    \addplot + [nodes near coords, fill=red!60]         table    [y=DocTer, meta=Tool, x expr=\coordindex] {\testdata};
\legend{FreeFuzz, 
DeepRel, 
DocTer,
}
\end{axis}
\end{tikzpicture}
}
\vspace{-0.1in}
  \caption{Distribution of detected bugs by the three fuzzers.}
  \label{fig:detectedbugdist}
\end{figure}

\subsection{RQ4: Root Causes for Missing Bugs} 
\label{sec:answer_rq4}


\textbf{Approach:} To address this RQ, we first thoroughly analyzed the implementation and the documentation of DL fuzzers to understand how they perform fuzzing based on two broad aspects, i.e., mutation strategies and oracles. We then summarized the issues that were missed by the fuzzers and created taxonomies of root causes. In terms of mutation strategies, we analyzed the root causes of missing bugs in FreeFuzz and DeepRel together since they share the same mutation strategy.

\textbf{Result:} Table \ref{tbl:rootcause} shows the summary of the root causes of missing bugs. The table has six columns. The first column indicates the broad root cause category of missing bugs, while the second column shows the corresponding breakdowns for each broad category. The third column, scope, shows the scope of root causes. Within scope means the fuzzers are able to model the root cause, while, out of scope refers to situations where the root cause category cannot be adequately modeled or addressed by the fuzzers. We summarized the root causes into four high-level categories, i.e., \textit{Randomness}, \textit{API Behaviors Related}, \textit{Oracle related}, and \textit{Boundary related} with the corresponding breakdowns. We categorized the rest, which does not belong to any other category, as \textit{Others}. 

\begin{table*}[t!]
\centering
\caption{Summary of root causes of missing bugs of the three DL fuzzers.}
\vspace{-0.1in}
 \begin{adjustbox}{width=0.8\textwidth}
\begin{tabular}{lllccc}
\toprule
Root Cause Categories & Breakdown                                       & Scope       & DocTer & FreeFuzz \& DeepRel & Total \\
\midrule
Randomness            & Specific Input Combinations                     & Within Scope & 16     & 30                  & 46    \\
\multirow{8}{*}{}     & Others                                          & Within Scope & 16     & 18                  & 34    \\
                      & Zero Dimensional Tensor                         & Within Scope & 7      & 6                   & 13    \\
                      & Empty input Tensor                              & Within Scope & -      & 9                   & 9     \\
                      & Tensor With Specific Data Type                  & Within Scope & 2      & 4                   & 6     \\
                      & Zero Input Argument                             & Within Scope & 2      & 3                   & 5     \\
                      & Multidimensional Tensors                        & Within Scope & 1      & 1                   & 2     \\
                      & Negative Integer                                & Within Scope & 1      & 1                   & 2     \\     
                      & Argument With Specific Data Type                & Within Scope & -      & 2                   & 2     \\
                      \midrule
API Behaviors Related & Lack of Testing API Use Chains                  & Out of Scope & 17     & 23                  & 40    \\
\multirow{2}{*}{}     & Lack of Modeling Complex API Context            & Out of Scope & 28     & 7                   & 35    \\
                      & Others                                          & Out of Scope & 14     & 11                  & 25    \\
                      \midrule
Oracle Related        & Lack of Capturing Inconsistent Results          & Out of Scope & 29     & 43                  & 72    \\
\multirow{2}{*}{}     & Lack of Capturing Execution Time                & Out of Scope & 7      & 9                   & 16    \\
                      & Lack of Capturing High Memory Usage             & Out of Scope & 5      & 5                   & 10    \\
                      \midrule
Boundary Related      & Lack of Generating Very large Integer           & Out of Scope & 12     & 11                  & 23    \\
\multirow{8}{*}{}     & Lack of Generating NaN Input                    & Out of Scope & 3      & 3                   & 6     \\
                      & Named Tensors                                   & Within Scope & 2      & 2                   & 4     \\
                      & Non Scalar Input                                & Within Scope & 2      & 1                   & 3     \\
                      & Lack of Generating  Contiguous Permuted Tensors & Out of Scope & -      & 3                   & 3     \\
                      & Nested List of Tensors                          & Within Scope & 1      & 1                   & 2     \\
                      & Lack of Modeling Non ASCI Input                 & Out of Scope & 2      & -                   & 2     \\
                      & Lack of Feeding Large Tensor                    & Within Scope & -      & 2                   & 2     \\
                      & Invalid Input String                            & Within Scope & 1      & -                   & 1     \\
                      \midrule
Others                & -                                               & Out of Scope & -      & 4                   & 4     \\
\bottomrule
\end{tabular}
\end{adjustbox}
\label{tbl:rootcause}
\end{table*}

\subsubsection{Randomness}
The bugs in this category arise from the inherent randomness of the employed fuzzers. In fact, the studied fuzzers have the ability to generate inputs that fall into the subcategories listed in Table \ref{tbl:rootcause}. However, this random behavior can pose a significant challenge in fuzz testing and may even be seen as a potential threat to the effectiveness of DL fuzz testing tools. In the following paragraphs, we explain the subcategories in detail. Please note that due to space limitations, we only elaborate on the top subcategory.

\noindent \textbf{Specific Input Combinations:}
Bugs in this category are triggered when specific inputs are fed into the DL APIs. For example, in this bug\footnote{https://github.com/tensorflow/tensorflow/security/advisories/GHSA-6gmv-pjp9-p8w8}, the API \textit{tf.raw\_ops.ReverseSequence} there is a mismatch between \textit{batch\_dim} and the rank of the input tensor. 

\subsubsection{API Behaviors Related}
Bugs in DL libraries can emerge from both backend implementation and front-end use of DL APIs. Bugs in the backend might be caused by flaws in the implementation of the library's algorithms, data structures (more specifically, tensors), or calculations (tensor level operations) \cite{harzevili2022characterizing}. These flaws can cause inaccurate results, memory leaks, and crashes. Misuse of the DL APIs might result in issues on the front end, such as erroneous parameter settings, mismatched data formats, or inaccurate input data. These issues might cause erroneous output or unexpected behavior from the DL models~\cite{harzevili2022characterizing}. These bugs are difficult to identify, reproduce, and fix due to their intricate nature. There are two subcategories of this category, which are as follows.

\vspace{4pt}
\noindent \textbf{Lack of Modeling Complex API Context:}
We find that some of the bugs require certain complex run-time contexts to be reproduced, while none of the DL fuzzers can provide even detailed API input specifications. Figure~\ref{fig:issue1} depicts an example of \textit{Internal Assertion Failure} in the PyTorch library\footnote{https://github.com/pytorch/pytorch/issues/77167}. In this bug, the DL API \textit{nn.functional.pad} is vulnerable to boolean values when used under the PyTorch just-in-time compiler. The reason for missing this bug by the three fuzzers is that they are not able to generate a test case with a call \textit{nn.functional.pad} under the context of PyTorch's just-in-time compiler. These fuzzers do not have fuzzing operators to model just-in-time related code patterns inside a test case.

\begin{figure}[t!]
\centering
    \begin{lstlisting} 
import torch
from torch import nn

class MyModule(nn.Module):
    def forward(self, inputs):
        return nn.functional.pad(
            inputs, (0, inputs.size(1) + 1), 
            value=False # works if value=0
        )
torch.jit.trace_module(MyModule(), 
{"forward": torch.zeros(3, 4)})
    \end{lstlisting} 
  \caption{An example of \textit{Inernal Assert Failure} in PyTorch library occurs when \textit{nn.functional.pad} is used inside a class module that has a forward method which is compiled using PyTorch just-in-time compiler.}
\label{fig:issue1}
\end{figure}



\vspace{4pt}
\noindent \textbf{Lack of Testing API Use Chains:} 
This root cause relates to bugs that are caused by multiple APIs being called together. While these fuzzers are designed to generate test cases by creating single usages for a specific API under test, Because of this intrinsic gap, these fuzzers cannot generate API usage chains, thus missing out on detecting bugs.

\subsubsection{Oracle Related} When testing software, the \textit{Oracle Problem} refers to the challenge of determining the correct output for a given input \cite{dou2023detecting, barr2014oracle}. It is especially problematic in the case of fuzzing since fuzzers generate random or semi-random inputs and must assess whether the output produced by the program under test is valid or not. Test oracles are also essential in DL library testing. The accuracy of the library's output is critical to ensuring that the programs that utilize it function as intended. It might be difficult to verify whether the library's output is valid or not without dependable test oracles. In this study, we summarized two subcategories that are related to oracles.
\vspace{4pt}

\noindent \textbf{Lack of Checking Logic Bugs:} 
In DL libraries, a logical bug is a defect in the design or implementation of the backend that leads it to give inaccurate or unexpected results despite the code running without error. Often, these vulnerabilities are difficult to detect since they do not trigger a crash or error message but instead give results that are inconsistent with the desired outcome~\cite{liang2022detecting}. Because of the nature gap between this type of bug and the intrinsic design choices of these DL fuzzers, i.e., their inability to infer logic oracles from code, these DL fuzzers miss detecting logic bugs.

\vspace{4pt}
\noindent \textbf{Lack of Checking Memory Leak:}  
In DL libraries, A memory leak happens when the software fails to release memory that it no longer requires, causing the program to use more and more memory over time \cite{harzevili2022characterizing}. This can ultimately cause the software to run out of memory, crash, or cause other system problems. We observed several instances of \textit{Memory Leak} in PyTorch and TensorFlow libraries in our curated benchmark dataset. For example, in this \textit{Memory Leak} example from the Pytorch library\footnote{https://github.com/pytorch/pytorch/issues/58109}, a memory leak is due to calling \textit{torch.jit.trace}. The log message in the bug description shows high memory usage as the number of iterations increases. Unfortunately, none of the employed fuzzers in this paper are able to detect \textit{Memory Leak} due to their oracle weaknesses. As mentioned earlier, DocTer is limited to crash oracles, and it cannot capture memory leaks since it only searches for specific crash bugs in the stack trace of the running generated test cases. Even though FreeFuzz and DeepRel are equipped with more oracles, they are still incapable of detecting high memory usage and memory leaks when the generated test cases are running.

\vspace{4pt}
\subsubsection{Boundary Related} 
Our curated dataset shows that in DL libraries, the use of boundary input values is one of the primary root causes of bugs. This is consistent with findings from the DocTer study \cite{xie2022docter} and previous research such as \cite{guo2020audee, dutta2018testing}, which suggest that off-by-one errors are the root cause of this issue. A recent empirical study on DL security vulnerabilities \cite{harzevili2022characterizing} highlights that missing validation checks in the backend implementation of DL libraries are the main cause of security vulnerabilities. Typically, these validation checks are intended to prevent weaknesses that arise from edge cases or boundary inputs that are intentionally or unintentionally provided to high-level DL APIs. As shown in Table \ref{tbl:rootcause}, we find three major root causes under this category, which are as follows.

\vspace{4pt}
\noindent \textbf{Lack of Generating Very Large Integer:} Figure \ref{fig:boundaryissue1} shows an example of \textit{Crash} bugs in the TensorFlow library when calling \textit{tf.math.segment\_max} with large segment ids. In this issue, the backend implementation of \textit{tf.math.segment\_max} fails to do a validation check on large integer values coming to backend~\cite{harzevili2022characterizing}. In the backend, the large values cause an \textit{Integer Overflow}.
The three fuzzers examined are not able to detect bugs that are due to feeding large integer values to DL APIs. Specifically,
DocTer does not support large integer values as its input boundary generation rules only focus on five generic boundary values, including \textit{None}, \textit{zero}, \textit{zero dimension}, \textit{empty list}, and \textit{empty string}. FreeFuzz and DeepRel are also incapable of generating very large integer values and only support the following values for their mutate integer rule: \{$-1024, -16, -1, 0, 1, 16, 1024$\}. The largest generated value in FreeFuzz and DeepRel is $1024$ which is not big enough based on the large integer values we observed in our curated dataset.

\begin{figure}[t!]
\centering

    \begin{lstlisting} 
import tensorflow as tf
tf.math.segment_max(data=np.ones((1,10,1)), 
segment_ids=[1676240524292489355])
    \end{lstlisting} 
    \vspace{-0.1in}
  \caption{An example of \textit{Crash} in TensorFlow due to large segment id's in \textit{tf.math.segment\_max} as input parameters. }
\label{fig:boundaryissue1}
\end{figure}

\begin{figure}[t!]

\centering
\begin{lstlisting} 
import torch
def unique(x):
    return torch.unique(x, sorted=False,
    return_inverse=False, return_counts=True)
s = torch.tensor(0.).cuda()
x = torch.tensor(float('nan')).cuda()
print(unique(s))
print(unique(x)) # <- these two calls have 
different outputs
print(unique(x)) # <-
    \end{lstlisting}
    \vspace{-0.1in}
  \caption{A bug in PyTorch caused by feeding \textit{NaN} input
  which results in non-deterministic behavior on GPU. }
\label{fig:boundaryissue2}
\end{figure}


\vspace{4pt}
\noindent \textbf{Lack of Generating NaN Input:} 
Figure \ref{fig:boundaryissue2} shows an example of a bug in the PyTorch library\footnote{https://github.com/pytorch/pytorch/issues/76571} due to feeding \textit{NaN Input} to \textit{torch.unique}, which results in inconsistent results as the nondeterministic output on GPU.
We find that neither FreeFuzz nor DeepRel support \textit{NaN} as boundary values. DocTer supports \textit{None} as a boundary value while does not support \textit{NaN}. There are inherent differences between \textit{NaN} and \textit{None} in Python. The reason is that in numerical operations, \textit{NaN} can be considered a numerical value, while \textit{None} is a \textit{Type}.

\vspace{4pt}
\noindent \textbf{Lack of Generating Non-ASCII Input:}
Some of the bugs are triggered by feeding \textit{Non-ASCII Input}. Unfortunately, none of the three DL fuzzers are able to generate Non-ASCII characters for test case generation. Even though the frequency of such bugs in our curated dataset is low, nonascii values can be considered a threat to DL libraries.

\mybox{\textbf{Answer to RQ4:} There exist four main categories of root causes for missing bugs by the studied fuzzers. \textit{Randomness} highlights the random nature of fuzzers, leading them to overlook a large number of bugs. \textit{API Behaviors Related} points to the lack of context information modeling in the generated test cases. \textit{Oracle Related} reveals the limitations of the studied fuzzers in detecting non-crash bugs. Lastly, the \textit{Boundary Related} category addresses the root cause of missing bugs arising from the lack of modeling edge cases.}
\section{Recommendations}
\label{sec:discussion}

Our study reveals several interesting findings that can serve as practical guidelines for both industry and academic communities to improve the development of DL fuzzers.

\noindent \textbf{Wrap test cases with API context mined from open source:}
For the purpose of improving the fuzzers' detection performance, we recommend future work toward improving these DL fuzzers take into account the context of the API in test case generation. There are many existing studies that mine API usage scenarios to extract useful context information for specific tasks, e.g., API misuse detection~\cite{ren2020api} and API knowledge graph construction~\cite{li2018improving}. It is feasible to mine API context information from code examples on Stack Overflow or Github. 


\noindent \textbf{Wrap test cases with API call chain:}
We recommend that further improvements to the studied fuzzers leverage API call chains to improve detection performance because certain bugs are only detectable when several APIs are invoked simultaneously. For example, one can employ API usage sequence/pattern mining techniques to extract DL API usage patterns and reuse them for modeling API call chains during test case generation ~\cite{zhong2009mapo, wang2013mining, wei2022api}.

\noindent \textbf{Expanding the input space with adding critical boundary values:}
The studied DL fuzzers are not able to generate some boundary values that are critical to finding edge cases in the backend implementation of DL libraries. To be able to find the edge cases, the studied fuzzers need to define new mutation operators. For example, the fuzzer can iterate over all input parameters and detect if the current input is an integer or not. If the input is an integer, the fuzzer can replace the existing values with a \textit{Large Integer Number}. 



\noindent \textbf{Expanding oracles with adding memory leak checks:}
As reported in \cite{harzevili2022characterizing}, \textit{Memory Leak} is a significant threat to DL libraries. In this study, we also find that \textit{Memory Leak} is also a prevalent bug in our curated dataset, as shown in Figure \ref{fig:bugdist}. While the studied three DL fuzzers are incapable of detecting \textit{Memory Leak} since their oracles are designed to identify crash and logical bugs only. One possible solution is to utilize memory profilers or memory checkers~\cite{nethercote2004dynamic, nethercote2007valgrind, bond2007tracking} in combination with the generated test cases. Specifically, after generating and executing a test case, a memory checker can be executed simultaneously to detect high memory usage. Once significant memory usage is detected, the profiler can signal the fuzzer to terminate the test case and consider it a potential bug. Although the operating system itself can indicate the termination of a test case due to memory bugs, external memory checkers are required since not all \textit{Memory Leak} bugs can cause the test cases to halt during their execution. 


\section{Improving DL Fuzzers with A Simple Corner Case Generator}
\label{sec:corner}

\begin{figure}[t!]
    \centering
    \resizebox{!}{0.2\columnwidth}{\includegraphics{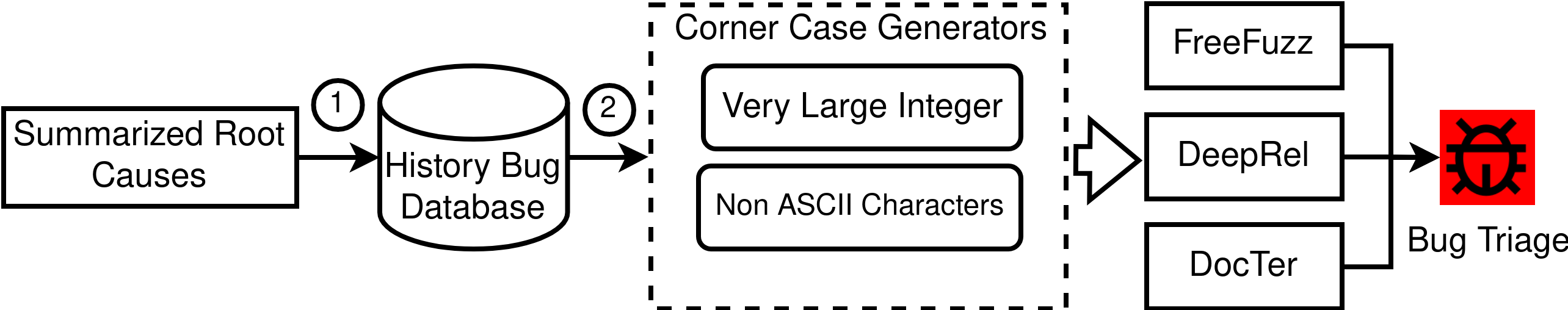}}
    \vspace{-0.1in}
    \caption{overview of corner case generator}
    \label{fig:cornercase}
\end{figure}

In Section~\ref{sec:discussion}, we recommend a set of guidelines to improve the fuzzers. As a proof of concept, in this section, we propose a corner case generator as an extension to the existing fuzzers, which efficiently encompasses multiple boundary values and DL-specific data types. This lightweight generator has the capability to produce test cases to identify bugs that might go unnoticed by the examined three DL fuzzers. The corner case is aimed at producing critical inputs that often elude conventional testing techniques.  
The proposed corner case generator, depicted in Figure~\ref{fig:cornercase}, comprises two major phases.

\noindent \textbf{Sample Corner Case Extraction}: using the collected DL bugs in Section~\ref{sec:step3}, we started to extract real-world malicious values for each specific corner case. For instance, when handling the scenario of \textit{Generating Very Large Integer Value}, we collect values such as \textit{[8968073515812833920, -1250999896764, 2**31-1, 100000000, -1e+38, 1676240524292489355, 36028797018963968]}. Additionally, for \textit{Non-ASCII Characters}, we instruct the corner case generator to mutate preemptive data type values using \textit{0x110000}.

\noindent \textbf{Corner Case Value Injection}: In this phase, the corner case generator applies mutations to the values whenever they encounter an input argument with data types such as integer, float, or string values. The generators use these data types as the basis for identifying specific areas where mutations can be introduced, thereby exploring a broader range of inputs for potential corner cases. {Mutation involves making alterations to data types and values using predefined rules in these fuzzers. For example, it includes transforming a small integer into a significantly larger integer value.}

Table~\ref{tbl:cornerdetections} presents the results of newly detected bugs by the three DL fuzzers integrated with our corner case generator. The table outlines the {type of corner inputs} and provides the corresponding count of detected bugs. When integrating our corner case generator as an extension of the FreeFuzz and DeepRel tools, they can identify 12 bugs triggered by feeding \textit{Very Large Integers} values for each tool. Three bugs can only be detected by FreeFuzz and DeepRel, but not by DocTer, and vice versa. Regarding \textit{NoN-ASCII Characters}, DocTer could detect two bugs while FreeFuzz and DeepRel could not detect them. Notably, among all the detected bugs, there were nine common bugs that were identified by all three fuzzers,  which highlights the corner case generator's effectiveness in detecting common bugs despite variations in the inherent nature of the fuzzers' implementations.




 


\begin{table}[t!]
\centering
\caption{Newly detected bugs by of DL fuzzers integrated with our corner case generator}
\vspace{-0.1in}
\begin{adjustbox}{width=0.45\textwidth}
\begin{tabular}{llc}
\toprule
Extended Fuzzers                      & Type of corner inputs                 & \# of bugs (\# unique)  \\
\midrule
\multirow{1}{*}{FreeFuzz/DeepRel} & Very Large Integer          & 12 (3)         \\
                          & Non ASCII Characters        & 0           \\
                          \hline
\multirow{1}{*}{DocTer}   & Very Large Integer          & 12 (3)         \\
                          & Non ASCII Characters        & 2           \\
                          \bottomrule
\end{tabular}
\end{adjustbox}
\label{tbl:cornerdetections}
\end{table}

\section{Related Work}
\label{sec:related}


In recent years, benchmarking studies in software engineering have attracted significant attention to assessing the efficiency and shortcoming of various software engineering methodologies, tools, or procedures in domains including but not limited to static bug detection~\cite{habib2018many, tomassi2018bugs, rutar2004comparison, zitser2004testing, chatzieleftheriou2011test, tomassi2021real}, reviewer recommendation system~\cite{gauthier2021historical}, API recommendation~\cite{peng2022revisiting, wei2022api}, and automatic test case generation~\cite{wang2021automatic}. Also, there have been multiple studies on Benchmarking fuzz testing tools on general software ~\cite{ hazimeh2020magma, metzman2021fuzzbench, natella2021profuzzbench, bohme2022reliability}. The reason behind multiple studies benchmarking fuzz testing tools is twofold. First, the evaluation of fuzzers is challenging since there is no solid metric to compare fuzzers. Crash oracles are the most straightforward technique, though they suffer from deduplication problem~\cite{jiang2021igor, klees2018evaluating}. Second, the lack of a benchmark dataset results in an unreliable comparison of fuzzers, which may introduce bias in the results.
Magma~\cite{hazimeh2020magma} is one of the first steps towards benchmarking fuzzers for general software to tackle the aforementioned issues. It created a benchmark dataset that contains real bugs collected from general software to enable uniform fuzzer evaluation and comparison. Magma allows for the realistic evaluation of fuzzers against a broad set of targets and enables the collection of bug-centric performance metrics independent of the fuzzer. 
Metzman et al.~\cite{metzman2021fuzzbench} introduced FuzzBench as a free, open-source platform for fuzzer evaluation. 
Natella et al.~\cite{natella2021profuzzbench} proposed a benchmark for stateful fuzzing of network protocols called ProFuzzBench. The benchmark consists of open-source programs that implement major network protocols, and tools for automating fuzzing experiments. 

Unlike the previous line of research in which benchmarking fuzzers on general software has received significant attention, DL fuzzers \cite{xie2022docter, wei2022free, deng2022fuzzing} are a very new area of research, hence there have not been a lot of benchmarking studies on them yet. Furthermore, benchmarking studies necessitate a large and carefully curated dataset of bugs for evaluation~\cite{hazimeh2020magma}, which is difficult to obtain for DL fuzzers. In this study, we take the first step toward benchmarking DL fuzzers. To accomplish this, we use a curated dataset of real-world bugs, which includes more bugs than previous studies, as we collected 627 bugs while existing studies have used 268 known bugs~\cite{bohme2022reliability}. We run three recently introduced DL fuzzers, including DocTer~\cite{xie2022docter}, FreeFuzz~\cite{wei2022free}, and DeepRel~\cite{deng2022fuzzing} on TensorFlow and PyTorch libraries with more than 4.3 million lines of code. The studied fuzzers could generate 161,417 test cases on the TensorFlow and PyTorch libraries. Based on the obtained results, our study performs an in-depth root-cause analysis of why the studied fuzzers miss real-world bugs. We then categorized the root causes into high-level and fine-grained root causes, which we believe are significant in improving the studied fuzzers. Based on the root cause analysis, our study provides a set of solid recommendations to improve the fuzzers in terms of detection effectiveness. 

\section{Threats to validity}
\label{sec:threats}
\noindent \textbf{Internal Validity:} In this section of the paper, we elaborate on the possible threats to the validity of our findings by benchmarking DL fuzzers. In order to avoid internal validity regarding running the studied fuzzers, we followed the original papers of the studied fuzzers to set up and run the fuzzers on the PyTorch and TensorFlow libraries. Regarding our corner case generator, the rationale behind choosing to implement this extension is that it is the easiest and most straightforward extension of the existing fuzzers. We take this extension as an example, and in future directions of this paper, we plan to implement more extensions for the fuzzers to improve their effectiveness at bug detection.

\noindent \textbf{External validity:} Regarding our data collection and classification, three authors manually checked the collected data in two rounds and resolved the possible disagreements. To guard against external validity, we run the fuzzers on two widely used DL libraries with 4.3 million lines of code. We also collect a significant number of real-world bugs (627 bugs) from public and open source domains such as the GitHub repository of the studied DL libraries as well as the CVE website and TensorFlow security advisories.

\section{Conclusion}
\label{sec:conclusion}
In this work, we conduct the first empirical study to evaluate state-of-the-art DL fuzzers, i.e., DocTer, FreeFuzz, and DeepRel. Specifically, we first collected a curated dataset of real-world DL bugs, we then select three state- DL fuzzers (DocTer, FreeFuzz, and DeepRel) that were tested on TensorFlow and PyTorch libraries, which have more than 4.3M lines of code and generated 161,417 test cases. We conducted an in-depth root-cause analysis of why the studied fuzzers miss real-world bugs based on the obtained results. We then categorized the root causes into high-level and fine-grained root causes that are significant in improving the studied fuzzers. Based on the root cause analysis, we provided solid recommendations to improve the fuzzers' detection effectiveness. As a proof of concept, we also proposed a corner case generator as an extension of the three fuzzers which help them detected 17 more bugs that were overlooked by the original fuzzers. 

\bibliographystyle{ACM-Reference-Format}
\bibliography{sample-base}

\end{document}